# Strength of Cu-TiN and Al-TiN interfaces from first-principles

S.K. Yadav[1], R. Ramprasad[2], J. Wang[1], A. Misra[3], X.-Y. Liu[1]

[1]Materials Science and Technology Division, MST-8, Los Alamos National Laboratory, Los Alamos, New Mexico 87545, USA

[2]Materials Science and Engineering, University of Connecticut, Storrs, Connecticut 06269, USA

[3]Materials Physics and Applications Division, MPA-CINT, Los Alamos National Laboratory, Los Alamos, New Mexico 87545, USA

**Abstract:**

Using density functional theory (DFT) based first principles calculations, we show that the preferred interfacial plane orientation relationship is determined by the strength of bonding at the interface. The thermodynamic stability, and the ideal tensile and shear strengths of Cu/TiN and Al/TiN interfaces are calculated. While there is a strong orientation relation (OR) preference for Al/TiN interface, there is no OR preference for Cu/TiN interface. Both the ideal tensile and shear strengths of Cu/TiN interfaces are lower than those of bulk Cu and TiN, suggesting such interfaces are weaker than their bulk components. By comparison, the ideal strengths of Al/TiN interface are comparable to the constituents in the bulk form. Such contrasting interfaces can be a test-bed for studying the role of interfaces in determining the mechanical behavior of the nanolayered structures.

## 1. Introduction

Metal-ceramic interfaces are of key importance in many applications including nanoelectronics, sensors, communication devices, composites, and catalysis [1]. Multilayered nanocomposites composed of alternating metal and ceramic layers have been actively explored experimentally. Such composites hold promise for extraordinary mechanical properties, and could lead to ductile, yet strong, materials [2-4]. At the nanometer length scales, the interfacial area per unit volume is significantly increased and thus, the bulk mechanical properties of the multilayers are dominated by interfaces [5]. We showed in our previous work that different interface chemistries might lead to contrasting ideal shear strength behaviors at the Al-TiN interface [6], thereby adding to the considerably vast literature on this topic [7-10].

Given the fact that interfaces can play a critical role in determining the mechanical behavior of nanolayered structures, a variation (or tuning) of interface bonding can sometimes serve a great purpose, from the perspective of controlling mechanical properties [11]. In this paper, we report a theoretical study of two different types of metal/ceramic interfaces, Al/TiN and Cu/TiN through first-principles DFT modeling. The DFT results suggest that the Cu/TiN interface is extraordinary "weak" in shear, significantly weaker than either TiN or Cu in the bulk form. This is in great contrast to the interface chemistry dependent Al/TiN interfaces, which, if N (or Ti) terminated leads to a strong interface with its strength comparable to that of TiN (or Al). The concept of "weak" interfaces has shown to be pivotal in determining the interface barrier to slip transmission in non-coherent metallic multilayer structures, for example, Cu/Nb [12-13], unlike the case of coherent Cu/Ni [14] where coherency stress dominates. Similarly, we expect the "weak" interface in metal/ceramic multilayers may have important implication to the dislocation slips during plastic deformation processes in these materials.

This paper is organized as follows. We first lay out the computational methodology in the Methods section. The Results section commences with the thermodynamic stability of both Al/TiN and Cu/TiN interfaces by calculating the formation energy of these interfaces and the associated works of adhesion of these interfaces. Subsequently the DFT results on the ideal tensile strengths of the considered interfaces are demonstrated. Finally, the ideal shear strengths of the considered interfaces are presented, which shows the most interesting feature of this work. We then conclude with a Summary.

## 2. Methods

Our DFT calculations were performed using the Vienna *Ab initio* Simulation Package (VASP) [15,16]. The DFT calculations employed the Perdew, Burke, and Ernzerhof (PBE) [17] generalized gradient approximation (GGA) exchange-correlation functional and the projector-augmented wave (PAW) method [18]. For all calculations, a plane wave cutoff of 500 eV for the plane wave expansion of the wave functions were used to obtain highly accurate forces. A 12x12x12 and 7x7x7 Monkhorst-pack mesh for k-point sampling are required to calculate elastic constants of metals and TiN ceramic respectively. A 7x7x1 Monkhorst-Pack mesh for k-point sampling is used for all calculations involving slabs. For slab calculations, a dipole correction perpendicular to the interface is added [19]. Table 1 lists the DFT calculated and experimental values of lattice parameters, bulk modulus, and elastic constants of Al, and Cu in face centered cubic (FCC) and TiN in rock salt crystal structure. The agreement between the DFT values and the experimental data is excellent [20-22].

Table 1. Comparison of calculated and experimental values [20-22] of lattice parameters, bulk modulus, and elastic constants of Al, Cu, and TiN.

|  | Al | | Cu | | TiN | |
| --- | --- | --- | --- | --- | --- | --- |
|  | DFT | Exp.[20] | DFT | Exp.[21] | DFT | Exp.[22] |
| Lattice Parameter (Å) | 4.04 | 4.04 | 3.63 | 3.61 | 4.24 | 4.24 |
| Bulk Modulus (GPa) | 76 | 79 | 137 | 142 | 277 | 288 |
| $C_{11}$ (GPa) | 114 | 108 | 170 | 176 | 639 | 625 |
| $C_{12}$ (GPa) | 61 | 62 | 120 | 125 | 139 | 165 |
| $C_{44}$ (GPa) | 25 | 28 | 77 | 82 | 160 | 163 |

For both Al/TiN and Cu/TiN interfaces, a textured growth along <111> direction for Al or Cu, and TiN, was preferred under experimental growth conditions [23-25]. Following the experiments, Al(111)/TiN(111) and Cu(111)/TiN(111) interfaces were considered in our DFT models. The computational supercell is a surface slab model, which includes 12 atomic layers of TiN (6 layers of Ti and 6 layers of N) and 6 atomic layers of Al (or Cu) along <111> direction. Additionally, a vacuum space of at least 6 Å was imposed on both surface sides to avoid surface-surface interactions. It is assumed that 6 atomic layers of metal or 12 atomic layers of TiN are thick enough to avoid any significant interaction of free surfaces with the interface region.

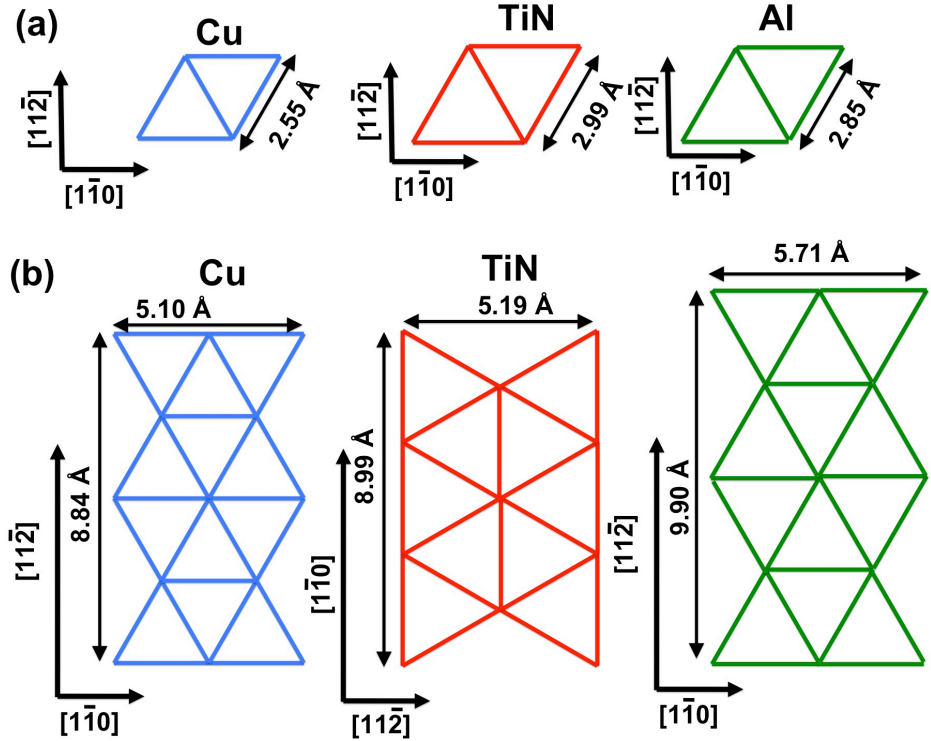

Fig. 1. Schematic diagram showing Cu (111) and Al (111) plane (a) matched and (b) rotated $90^0$ with respect to the underlying TiN (111) plane. Corners of triangles represent position of atoms.

The possible in-plane orientation relationships (OR) between metal (Al or Cu) and TiN ceramic are illustrated in Figure 1. Two in-plane OR between metal and TiN are considered: a) the lattice of metal matches with that of TiN along [1-10] and [11-2] directions (see Fig. 1a); b) the lattice of metal along [1-10] direction matches that of TiN along [11-2] direction (see Fig. 1b). In the paper, we refer to case a) as "*matched*" and case b) as "*rotated*" interface. In the *matched* interface Al or Cu atoms sit at FCC lattice positions with respect to the Ti atoms in the underlying TiN layers. In the *rotated* interface there is no one-to-one correspondence of the metal atoms with regard to either the Ti or N atoms in the underlying TiN layers. The misfit strains for both the "*matched*" and "*rotated*" cases are listed in Table 2. To form a coherent interface of Al on the TiN (111) plane, lattice strains of 4.8% and 11.2% are required for *matched* and *rotated* lattice orientations, respectively. From lattice mismatch alone, it is intuitive that Al/TiN interface would prefer the *matched* interface orientation, as observed in experiment [23]. To form coherent interface of Cu on TiN (111) plane, lattice strains of 16.1% and 1.7% are required for *matched* and *rotated* lattice orientations, respectively. In earlier experiments, both *matched* and *rotated* interfaces of Cu(111)/TiN(111) were observed [24], although based on lattice misfit alone it is intuitive that Cu/TiN interface would prefer the *rotated* interface orientation. In our DFT simulations, we modeled the *matched* lattice orientations for Al/TiN, and both *matched* and *rotated* lattice orientations for Cu/TiN interfaces. In all cases, the metal is strained to match the in-plane lattice parameter of TiN to form coherent interfaces. This is reasonable since the elastic constants of metals considered here are much smaller than those of TiN.

Table 2. Lattice strain for *matched* and *rotated* metal/ceramic interfaces.

|  | Strain (%) | |
| --- | --- | --- |
|  | Cu/TiN | Al/TiN |
| "Matched" | 16.1 | 4.8 |
| "Rotated" | 1.7 | 11.2 |

**3. Results**

**3.1. Structural and thermodynamic properties of interfaces**

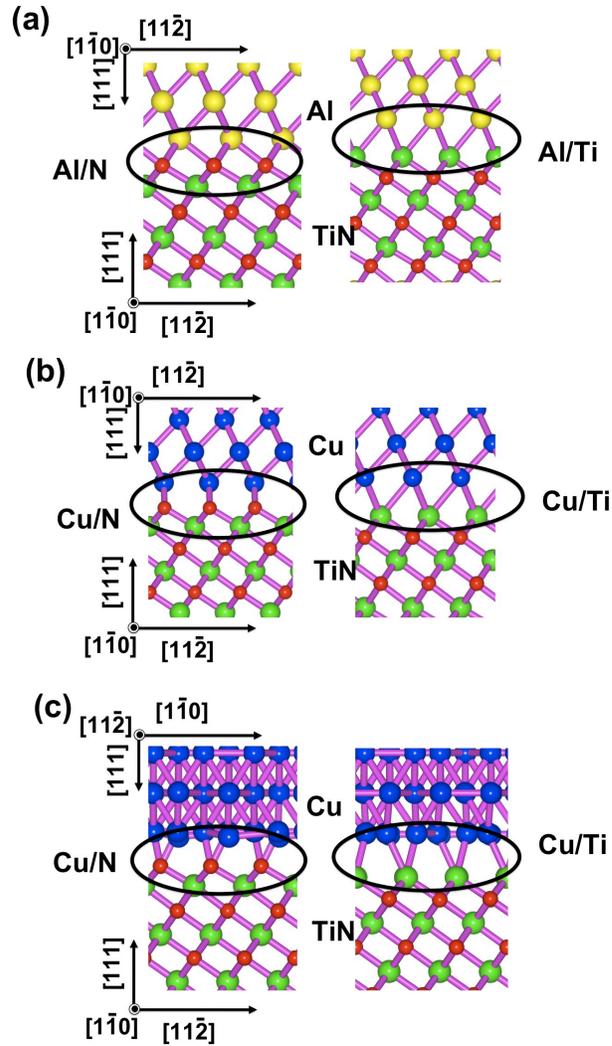

Fig. 2. Atomic structure of (a) *matched* Al/TiN interface with N and Ti termination, (b) *matched* Cu/TiN interface with N and Ti termination, (c) *rotated* Cu/TiN interface with N and Ti termination.

In Fig. 2a-c, the relaxed interface structures of the *matched* Al/TiN interface with N and Ti termination, the *matched* Cu/TiN interface with N and Ti termination, and the *rotated* Cu/TiN interface with N and Ti termination are shown. In the *matched* interface of Al/TiN, there are three possible positions for Al with respect to the underlying Ti in TiN: FCC, HCP (hexagonal close packed) and OT (on top) sites [23]. The most preferred position of Al with respect to the underlying Ti is FCC position, irrespective of Ti or N termination at the interface. In N terminated interface, each Al atom binds with three N atoms with an average bond length of 2.19 Å, while in the Ti terminated interface, each Al atom binds with three Ti atoms with an average bond length of 2.70 Å.

In the *matched* interface of Cu/TiN, the most preferred position of Cu is not FCC but HCP in the case of the N terminated interface. Since Cu does not have any affinity for N, the number of bonds between each Cu atom with N at the interface is reduced to two with an average bond length of 2.07Å. However, in the case of Ti terminated interface, the most preferred position of Cu is FCC. In this case, the HCP position is lower in energy by 0.73 J/m$^2$ compared to FCC position. Cu forms metallic bond with Ti, and in Ti terminated interface, the averaged bond length is also 2.70 Å. For the *rotated* interface of Cu/TiN, there is no one-to-one correspondence for the metal atoms with regard to either the Ti or N atoms in the underlying TiN. There is only one bond between Cu and N with an average bond length of 1.97 Å and one bond between Cu and Ti with an average bond length of 2.51 Å.

Using a coherent interface model, the formation energy of the interface is calculated. As interfaces are polar (TiN terminating in either N or Ti), the interface formation energy is defined as a function of the chemical potential of N,

$$E_{Interface} = \frac{E^{SC} - n_M E_M^{strain} - n_{Ti} E_{TiN} - (n_N - n_{Ti})\Delta\mu_{N_2}}{Area} \qquad \text{Eq. (1)}$$

where $E^{SC}$ is the DFT calculated total energy of the supercell. $n_M$, $n_{Ti}$, and $n_N$ are number of Al or Cu atoms, Ti atoms and N atoms, respectively. $\Delta\mu_{N_2}$ is the chemical potential of nitrogen. The chemical potential of nitrogen is constrained by the formation energy of TiN $\Delta\mu_{N_2} + \Delta\mu_{Ti} \leq F_{TiN}$, where $F_{TiN}$ is formation energy of TiN. Our DFT calculated formation energy of TiN is -3.5 eV, hence $\Delta\mu_{N_2}$ and $\Delta\mu_{Ti}$ can vary from 0 to -3.5 eV. "Area" is the total surface area of the interface in the supercell. $E_{TiN}$ is calculated bulk equilibrium DFT energy per TiN. $E_M^{strain}$ is the calculated DFT energy per atom of bulk Al or Cu with the same in-plane strain as induced when forming coherent interfaces in the supercell.

The chemical potential of $N_2$ depends on the temperature and pressure of $N_2$ during growth of the film. A useful definition of the chemical potential of $N_2$, is [26]

$$\mu_{N_2}(T, P_{N_2}) = E_{N_2}^{DFT} + \Delta\mu_{N_2}(T, P_0) \qquad \text{Eq. (2)}$$

$$\mu_{N_2}(T, P_{N_2}) = E_{N_2}^{DFT} + E_{N_2}^{ZPE} + \Delta\mu'_{N_2}(T, P_0) + kT\ln(P_{N_2}/P_0), \qquad \text{Eq. (3)}$$

where $E_{N_2}^{DFT}$ and $E_{N_2}^{ZPE}$ are the DFT energy and the zero-point vibrational energy of an isolated $N_2$ molecule at 0 K. The second equality of the above equation define $\Delta\mu_{N_2}(T, P_0)$, and $\Delta\mu'_{N_2}(T, P_0)$ is the difference in the chemical potentials of $N_2$ at (0 K, $P_0$) and at (T, $P_0$), where $P_0$ is the reference pressure, taken generally to be 1 atm. Using standard expressions for the molecule partition functions for an ideal diatomic molecule gas [27, 28], one obtains

$$\mu'_{N_2}(T, P_0) = kT\ln\frac{P_0}{(2\pi mkT/h^2)^{1.5}kT} - kT\ln\frac{T}{2\theta_r} + kT\ln(1-e^{-\theta_v/T}) - kT\ln 1 \qquad \text{Eq. (4)}$$

The four terms on the right represent the translational, rotational, vibrational, and electronic contributions, respectively. $h$ and $m$ are the Planck's and the mass of a $N_2$ molecule, respectively. $\theta_r$ and $\theta_v$ are the characteristic rotational and vibrational temperatures, respectively. The factor 1 in the last term accounts for the fact that one $N_2$ molecule has one spin configuration. The procedure to calculate chemical potential is described in more details in Ref. [29]. Based on the experimental growth condition, with pressure ranging from $10^{-3}$ atm to $10^{-2}$ atm and the temperature ranging from 400 to 500 K, the calculated chemical potential range is from -0.9 to 1.2 eV.

The work of adhesion of interfaces, which is defined as the amount of energy required to separate the interface into two free surfaces, is yet another way to capture the energetics of interface. The work of adhesion is calculated as below,

$$E_{Ahesion} = E_{Interface} - E_{Metal}^{Surface} - E_{Ti/N}^{Surface} \qquad \text{Eq. (5)}$$

where $E_{Interface}$ is the interface formation energy calculated above, $E_{Metal}^{Surface}$ is formation energy of Cu or Al (111) surface, and $E_{Ti/N}^{Surface}$ is formation energy of TiN (111) surface with Ti or N termination. By this definition, a stronger interface will have a more negative work of adhesion value.

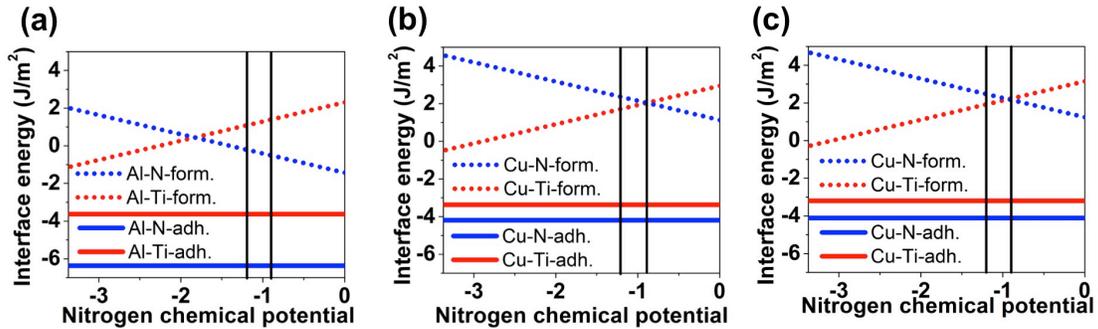

Fig. 3. Interface formation energy (form.) as a function of N chemical potential, and work of adhesion (adh.) of (a) Al/TiN *matched* interface (b) Cu/TiN *matched* interface (c) Cu/TiN rotated interface. Vertical lines show range of chemical potential of $N_2$ generally observed during growth of such interface. Metal-N indicates N terminated interface while metal-Ti indicates Ti terminated interface.

Figure 3 shows the DFT calculated interface formation energies as a function of chemical potential for both Ti and N terminations at the interface. In addition, the work of adhesion for Al/TiN and Cu/TiN interfaces are also shown. For Al/TiN interface, the N terminated interface has lower formation energy than Ti terminated interface for N chemical potential observed during growth. For the work of adhesion, N terminated interface is lower in energy due to the stronger bonding between Al and N atoms at the interface, compared to Al-Ti bonds at Ti terminated interface.

For Cu/TiN, both *matched* and *rotated* interfaces show similar behavior. In general, the interface formation energies in the Cu/TiN cases are substantially higher than the corresponding values in the Al/TiN cases. In fact, almost all the formation energy values are in the positive range in both *matched* and *rotated* interfaces, irrespective of N or Ti terminations at the interface. This indicates a substantially weak bonding between Cu and N at the interface. Experimentally, Cu forms only metastable bulk compounds with nitrogen [30-32]. In the experimental chemical potential range of N, both N terminated interface and Ti terminated interface can form.

### 3.3. Ideal tensile strength

The ideal strength, the highest achievable theoretical strength of a material, is the minimum stress needed to plastically deform an infinite dislocation-free crystal. An accurate estimate of the ideal strength is central to understanding the limits of mechanical strength of nanostructured materials such as multilayer films. In this paper, the ideal tensile and shear strengths of the "strong" Al/TiN interface and "weak" Cu/TiN interfaces are calculated.

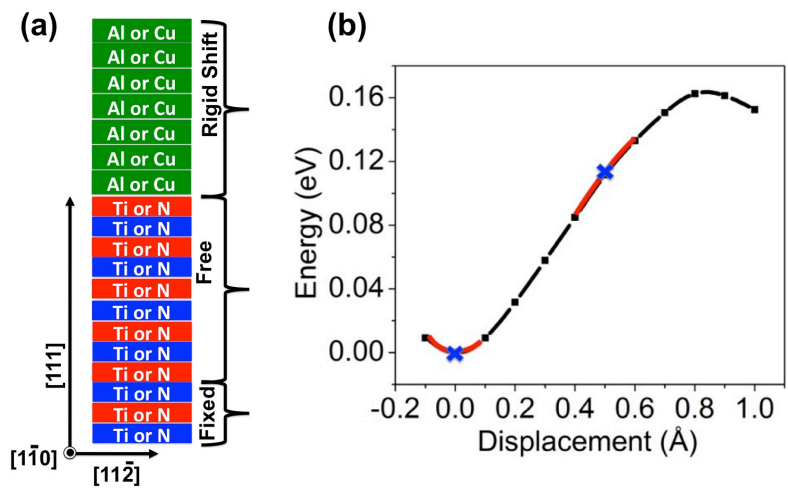

Fig. 4. (a) Schematic showing layers of Ti, N and Cu or Al that are fixed, free to relax and rigidly moved upward during tensile test of interface. (b) Schematic of curve fitting at each point by considering adjacent points.

For the ideal tensile strength simulations, a supercell containing 6 atomic layers of Al or Cu and 12 atomic layers of TiN (6 layers of Ti and N) with vacuum on either side was used. Six interfaces were considered, including the *matched* Al/TiN interface with N or Ti termination at the interface, the *matched* Cu/TiN interface with N or Ti termination at the interface, the *rotated* Cu/TiN interface with N or Ti termination at the interface. As schematically shown in Figure 4a, metal layers are rigidly shifted upwards with successive small displacements, in steps of 0.1 Å while keeping the bottom few layers of TiN partially fixed during the simulation. For each displacement, atoms in the fixed region are only allowed to relax within the (111) plane, and atoms in the free region are allowed to relax in all directions (see Figure 4a). As the supercell is perturbed from its equilibrium position, the energy of the supercell increases. The tensile stress at each level of displacement is calculated by taking the derivative of energy displacement curve at each point, and then divided by the area of the interface. The curve at each point is fitted to a third order polynomial by considering adjacent points, as schematically shown in Figure 4b. We would like to point out that fitting the curve at each point separately is a necessary step to obtain the accurate value of stress at each point. This is due to the dramatically changing curvature of energy vs displacement curve.

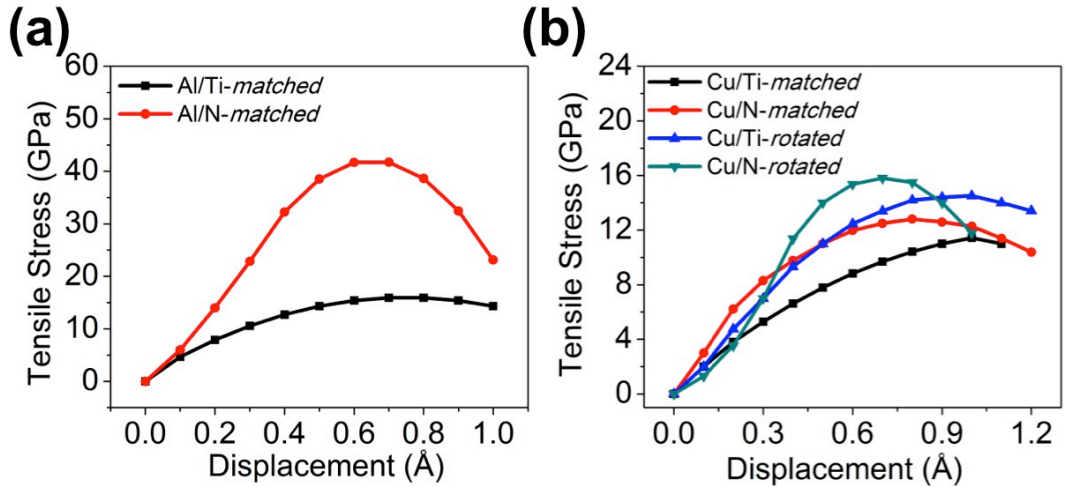

Fig.5. Stress-displacement curve during the tensile strength simulations as a function of displacement.

Figure 5 shows the plots of the stress-displacement curve during the tensile strength simulations. Initially the stress increases and then it reaches a maximum before mechanical failure, which gives the ideal tensile strength of the interface. Table 3 summarizes the ideal tensile strength for various cases considered. For the *matched* Al/TiN interface, the ideal tensile strengths obtained are 42 GPa with N termination at the interface, and 12 GPa with Ti termination at the interface. For both *matched* and *rotated* Cu/TiN interfaces, the ideal tensile strengths obtained are 15 GPa with N termination at the interface, and 13 GPa with Ti termination at the interface. To compare, the ideal tensile strengths of bulk crystal Cu, Al, and TiN along <111> direction from our separate DFT calculations are 19, 11, and 44 GPa, respectively. The ideal tensile strengths of the Cu/TiN interfaces are lower than that in bulk Cu, suggesting that the Cu/TiN interfaces are weaker than bulk Cu.

Table 3. Ideal tensile strength of the Al/TiN with N termination (Al/N) and Ti termination (Al/Ti) and Cu/TiN interface with N termination (Cu/N) and with Ti termination (Cu/Ti). Tensile strength along <111> for Cu, Al, and TiN are 19, 11, and 44 GPa.

|  | Cu/N (GPa) | Cu/Ti (GPa) |
|---|---|---|
| *Matched* | 15 | 13 |
| *Rotated* | 15 | 13 |
|  | Al/N (GPa) | Al/Ti (GPa) |
| *Matched* | 42 | 12 |

**3.3. Ideal shear strength**

To calculate the ideal shear strength, a series of incremental shear strains were applied to the suitably chosen supercell as depicted in Figure 6. For the *matched* Al/TiN and Cu/TiN interface, the shear strength was calculated along the <112> direction, as it was found earlier that for both Al and TiN, the ideal shear strength along this direction is lower than along other directions [6]. For Cu/TiN the *rotated* interface, the shear strength was calculated along both <110> and <112> directions of TiN. At the interface, atoms were allowed to relax in all directions except in the direction of shear displacement. [33-35] Similar to calculations for the tensile stress, the shear stress at each level of displacement is calculated by taking the derivative of energy displacement curve at each point and divided by the area of the interface.

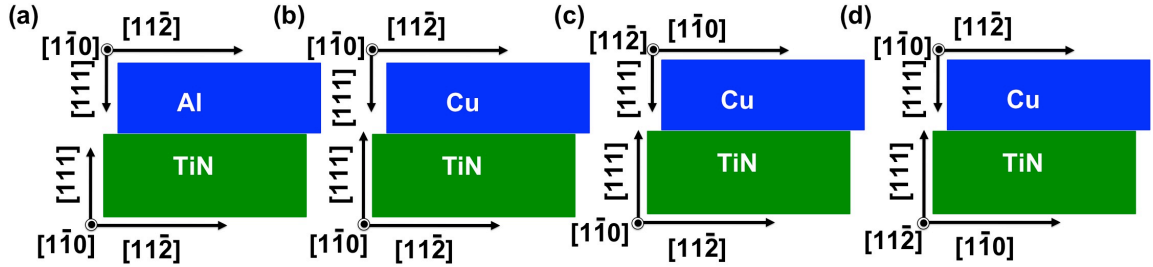

Fig. 6. Schematic showing direction of shear at *matched* (a) Al/TiN and (b) Cu/TiN interfaces and *rotated* (c) and (d) Cu/TiN interface.

Figure 7 shows the plots of the stress-displacement curve during the shear strength simulations as a function of displacement along the shear directions. Table 4 summarizes the ideal shear strength for various cases considered. Again, initially stress increases and then it reaches a maximum, which is taken as the ideal shear strength of the interface.

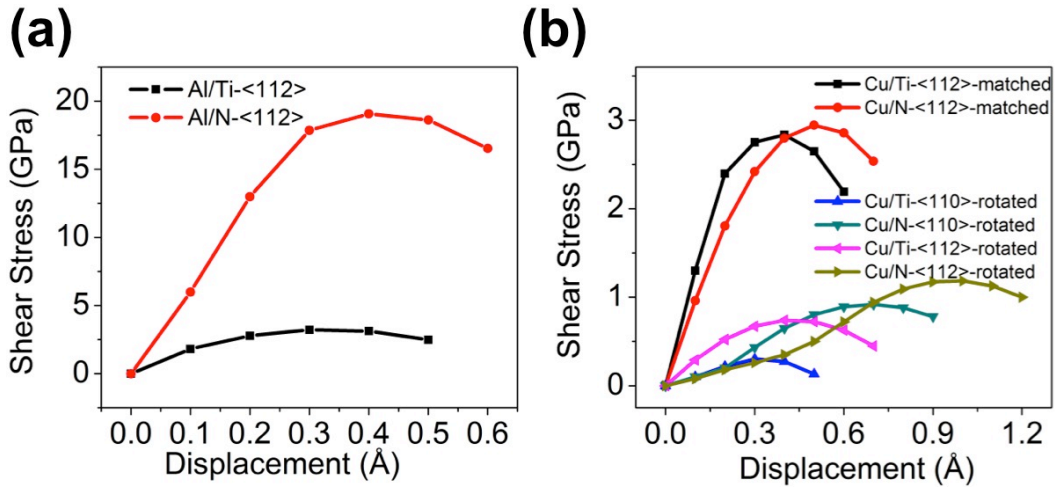

Fig.7. Stress-displacement curve during the shear strength simulations as a function of displacement along the shear directions.

For the *matched* Al/TiN interface, the ideal shear strengths along the <112> direction are 19 GPa with N termination at the interface, and 3.3 GPa with Ti termination at the interface. For the *matched* Cu/TiN interfaces, the ideal shear strength along the <112> direction is 2.8 GPa with N termination at the interface, and 2.9 GPa with Ti termination at the interface. For the *rotated* interface of Cu/TiN, the ideal shear strength along the <110> direction is 0.8 GPa with N termination at the interface, and 0.3 GPa with Ti termination at the interface; and along the <112> direction: 1.2 GPa with N termination at the interface, and 0.7 GPa with Ti termination at the interface. In general, the shear strengths at Cu/TiN interfaces are smaller than those in Al/TiN interfaces. There is a dramatic difference between the ideal shear strength of *matched* and *rotated* Cu/TiN interfaces. This may be due to the artifact associated with the straining of Cu to a larger extent than in the *matched* Cu/TiN interface case. Taking this consideration, we speculate that the value of the ideal shear strength calculated for the *rotated* Cu/TiN interface is more accurate. Our separate DFT simulations show that the lowest ideal shear strengths of bulk Cu, Al, and TiN are along the <112> direction on the (111) plane, with values of 3.0, 3.2, and 45.4 GPa, respectively. The ideal shear strengths of the Cu/TiN interface are generally lower than that of bulk Cu, especially true for the *rotated* Cu/TiN interfaces, where the ideal shear strengths are substantially lower.

Table 4: Ideal shear strength of Al/TiN and Cu/TiN interface. Direction of shear is defined with respect to TiN. Ideal shear strength of Cu, Al, and TiN along the <112> direction on the (111) plane is 3.0, 3.2, and 45.4 GPa, respectively.

|  | TiN Direction | Al/Ti (GPa) | Al/N (GPa) |
|---|---|---|---|

| Matched | <112> | 3.3 | 19.0 |
|---|---|---|---|
|  |  | Cu/Ti (GPa) | Cu/N (GPa) |
| Matched | <112> | 2.9 | 2.8 |
| Rotated | <110> | 0.3 | 0.8 |
|  | <112> | 0.7 | 1.2 |

**4. Summary**

To summarize, we performed accurate DFT based first-principles simulations of the two contrasting interfaces Al(111)/TiN(111) and Cu(111)/TiN(111). We show that in-plane orientation relation between metal/ceramic is not only determined by lattice mismatch but also by the strength of bonding at the interface. Due to the strong bonding between Al and N at interfaces, there is a strong orientation relation preference for the Al/TiN interface. For Cu/TiN interfaces, since the bonding between Cu and N at the interface is weak, there is no orientation relation preference for Cu/TiN interfaces. The thermodynamic stability, the ideal tensile and shear strengths of six types of Al/TiN and Cu/TiN interfaces are calculated. It is found that interface formation energies in the Cu/TiN cases are substantially higher than the corresponding values in the Al/TiN cases, with almost all the formation energy values in the positive range, irrespective of N or Ti terminations at the interface. Both the ideal tensile and shear strengths of Cu/TiN interfaces are smaller than those of bulk Cu and TiN, indicating that the interface is the weakest link in this system (unlike in Al/TiN interfaces). This study of prototypical Cu/TiN and Al/TiN interfaces underlines the power of first principles computations in the assessment of interfacial strengths, and the subsequent interface design of high-strength nanocomposites.

**Acknowledgements**

We would like to thank Nan Li for helpful discussions. This work was supported by the US Department of Energy, Office of Science, Office of Basic Energy Sciences. XYL also acknowledges partial support by the Los Alamos National Laboratory (LANL) Directed Research and Development Program. LANL is operated by Los Alamos National Security, LLC, for the National Nuclear Security Administration of the U.S. Department of Energy under Contract No. DE-AC52-06NA25396.